\documentstyle[11pt,aaspp4]{article}

\singlespace
\begin{document}

\title{Thermal Properties of Two-Dimensional 
       Advection Dominated Accretion Flow}

\author{Myeong-Gu Park}
%\author{Myeong-Gu Park\altaffilmark{1}}
\affil{Department of Astronomy and Atmospheric Sciences,
                 Kyungpook National University, Taegu 702-701, KOREA}
\author{Jeremiah P. Ostriker}
\affil{Princeton University Observatory, Princeton University,
       Princeton, NJ 08544}

%\altaffiltext{1}{also at Princeton University Observatory, 
%                 Princeton University, Princeton, NJ 08544}

\begin{abstract}
We study the thermal structure of the widely adopted two-dimensional
advection dominated accretion flow (ADAF) of Narayan \& Yi (1995a).
The critical radius for a given mass accretion rate, 
outside of which the optically thin hot 
solutions do not exist in the equatorial
plane, agrees with one-dimensional study. 
However, we find that,
even within the critical radius, 
there always exists a conical region of the flow,
around the pole, which cannot maintain 
the assumed high electron temperature,
regardless of the mass accretion rate, 
in the absence of radiative heating. 
This could lead to
torus-like advection inflow shape
since, in general, the ions too will cool down.
We also find that Compton preheating is generally important
and, if the radiative efficiency, defined as the luminosity output
divided by the mass accretion rate times the velocity of light
squared, is above $\sim 4\times10^{-3}$, the polar region of the flow
is preheated above the virial temperature by Compton heating
and it may result in time-dependent behaviour or outflow
while accretion continues in the equatorial plane.
Thus, under most relevant circumstances, ADAF solutions may
be expected to be accompanied by polar outflow winds.
While preheating instabilities exist in ADAF, as for
spherical flows, the former are to some extent protected
by their characteristically higher densities and
higher cooling rates, which reduce their susceptibility
to Compton driven overheating.
\end{abstract}


\section{Introduction}

Depending on the angular momentum it carries and how the 
angular momentum is dissipated, gas accretes onto compact objects
in various ways. If the gas contains very little angular momentum
or bulk flow, the flow becomes spherical 
(Hoyle \& Lyttleton 1939; Bondi \& Hoyle 1944;
Bondi 1952; Loeb \& Laor 1992; Foglizzo \& Ruffert 1997; 
Nio, Matsuda, \& Fukue 1998). If it has enough angular momentum
and if the angular momentum is efficiently
removed by some mechanism, the flow flattens to a disk shape 
(Pringle \& Rees 1972; Shakura \& Sunyaev 1973). 
Until recently, only the extreme of these two types of solutions 
have been studied (see Chakrabarti 1996a,b for history and unified
scheme of accretion solutions and Park \& Ostriker 1998 for review
on general accretion flow). 

Spherical accretion has the merit of being simple, and therefore
can be accurately calculated. It generally has the infall velocity
close to the free-fall value, and most of the gas energy generated
is lost into the hole in the case of accretion onto black holes,
making the radiation efficiency, $e$, of the accretion,
a self consistently calculable quantity depending primarily
on the entropy at large radii and the accretion rate.
The efficiency can be very low or significantly high 
(Shapiro 1973a,b, 1974; M\'esz\'aros 1975;
Park \& Ostriker 1989; Park 1990a,b). 
In the opposite limiting case
accretion proceeds in the form of thin disk
when the gas has enough angular momentum
and cools efficiently. It has a fixed
and high radiation efficiency, $e \sim 0.1$, and does not emit
high energy photons due to the low temperature of the gas.
Although the thin disk accretion has been successfully applied
to many astronomical sources (see Pringle 1981 and Frank, King,
\& Raine 1992 for reviews), its radiation spectrum is
incompatible with certain kinds of celestial sources
believed to be powered by accretion. 

There are also other types of accretion disk solutions 
beyond the cool thin disk. 
Shapiro, Lightman, \& Eardley (1976) showed that
geometrically and optically thin, high-temperature accretion disk can exist
in which ions and electrons are weakly coupled and
have different temperatures. Although
it was more successful in explaining high-energy sources like
Cyg X-1, it proved to be unstable on a thermal time scale
(Pringle 1976; Piran 1978; Park 1995). 
Seminal works on geometrically thick accretion disks appeared
a few years later (Jaroszy\'nski, Abramowicz, \& Paczy\'nski 1980; 
Paczy\'nski \& Wiita 1980). Abramowicz et al. (1988) further
combined these works with `$\alpha$-viscosity' model to find
a so called `slim disk' solution.
It is a geometrically and optically thick accretion disk solution
with a significant fraction of the gas energy being
transported via advection, but treated in one-dimensional framework.
These slim disk solutions exist only
when the mass accretion rate is near or above the Eddington mass 
accretion rate defined as
\begin{equation} 
   \dot M_{Edd} \equiv \frac{L_E}{c^2} 
                = 2.19\times 10^{-9}~M_{\sun}~{\hbox{yr}}^{-1}
\end{equation}
where $L_E$ is the Eddington luminosity and the numerical
value is for the pure hydrogen. (In the case of accretion onto
neutron stars or thin disk accretion, the Eddington accretion
rate has sometimes been defined as $e^{-1} L_E/c^2$ since $e$ is
almost a fixed value. However, in the type of accretion in which $e$
is not known a priori, this
definition would be confusing and misleading because
even when $\dot M \sim \dot M_{Edd}$, 
the luminosity from the accretion
can be much smaller than $L_E$. This choice of definition makes
$\dot M_{Edd}$ in this work 10 times smaller than $\dot M_{Edd}$ 
in NY3.) Slim disks have 
a radiation spectrum similar to that of the thin disks. 

Only recently, a new type of accretion solution has been found 
that stands somewhere between the spherical and thin disk accretion
(Narayan \& Yi 1994, 1995a,b, hereafter NY1, NY2, and 
NY3, respectively, and NY collectively; Abramowicz et al. 1995;
see Narayan, Mahadevan, \& Quataert 1998 for review and references).
These solutions are called `advection dominated accretion flow' (ADAF)
because most of the gas energy is advected with the flow and
ultimately into the hole due to relatively large infall
velocity. This type of accretion is possible at sufficiently
low mass accretion rate so that radiative cooling is inefficient
and ions can be kept near virial temperature.
The isodensity contours of self-similar
ADAF changes from that of a sphere to a somewhat flattened
spheroid, depending on the parameters chosen (NY2), but in
general the flow looks more like the spherical or spheroidal limit 
rather than the disk-like solutions.
It is in a way a reminiscence of ion-torus model (Rees et al. 1982).

ADAF has a number of distinct and desirable properties. 
First, like spherical accretion, 
the radiation efficiency can span a large range and is
generally much smaller than $\sim 0.1$. Second, the electron
temperature is much higher than that in thin disk accretion
and, therefore, can produce high-energy photons. Third, the
self-similar form of the solutions makes calculation
of various properties very easy (Spruit et al. 1987; NY1). 
These merits have led to
many successful applications of ADAF to various accretion
powered sources that are difficult to model by the standard
thin disk solutions: Sgr A$^\ast$
(Narayan, Yi, \& Mahadevan 1995; Mahadevan, Narayan, \& Krolik 1997;
Manmoto, Mineshige, \& Kusunose 1997; Narayan et al. 1998), 
NGC 4258 (Lasota et al. 1996), soft X-ray transient sources 
(Narayan, McClintock, \& Yi 1996; Narayan, Barret, \& McClintock 1997),
low-luminosity galactic nuclei (Di Matteo \& Fabian 1997; Mahadevan 1997),
torque-reversing X-ray pulsars (Yi \& Wheeler 1998), and
X-ray background (Yi \& Boughn 1998).

However, most elaborate works on ADAF have been based on
one-dimensional height-integrated equations (Narayan, Kato, \& Honma 1997;
Chen, Abramowicz, \& Lasota 1997;
Nakamura et al. 1997; Gammie \& Popham 1998; Popham \& Gammie 1998;
Chakrabarti 1996a) even though the desirable properties
depend explicitly on the two-dimensional characteristics of the flow.
The two-dimensional nature of ADAF has been explored only
in self-similar form (NY2; Xu \& Chen 1997)
or by numerical simulations of adiabatic, inviscid 
flow very close to the hole
(Molteni, Lanzafame, \& Chakrabarti 1994; Ryu et al. 1995;
Chen et al. 1997; Igumenshchev \& Beloborodov 1997). 
When treated as a height-integrated disk instead of two-dimensional flow, 
locally produced photons
are assumed to escape the flow without affecting
other other parts of the flow, which is only true when the flow
is truly disk-like. Also, parts of the thermal and dynamical
structure of the two-dimensional
flow can be widely different from those of the averaged disk,
under general accretion conditions, because of the flow structure
in the vertical direction.

So, in this work, we study the thermal properties of the
ADAF based on the two-dimensional self-similar flow solutions
of NY2. One important difference between this work and previous
studies is the allowance for the preheating of the flow 
at large radii by photons
produced at the inner, hotter part of the flow. When the luminosity
is high enough, this can substantially change the dynamics of the ADAF
leading, quite possibly, to some kind of time-dependent behaviour
or outflow.

\section{Thermal Properties of Advection Dominated Accretion Flows}

In generic accretion flow, the thermal state of the 
gas at a given point is determined by the balance between
various heating and cooling processes. 
Gas can be heated by $PdV$ work,
viscous dissipation, and the interaction with radiation. 
It cools mostly by emission of radiation. 
However, heating and cooling are not always
balanced, and the extra energy gain or loss will be stored as internal
energy and carried with the flow. This advection of internal
energy mostly acts as loss of gas energy at a fixed point 
in the accretion flow and sometimes is called advective cooling. 

The advection-dominated accretion flow solutions of NY
refer to a family of solutions where gas at a given point
is heated by viscous dissipation and cooled mostly by
the advection with negligible radiative cooling.
However, the definition of ``advective cooling'' in NY
contains the $PdV$ adiabatic heating term. 
So one should be careful when
advective cooling is used in thermal balance calculations. 
For example, gas of an adiabatic index,
$\gamma = 5/3$, and temperature profile, $T \propto r^{-1}$, 
has zero advective cooling because $PdV$ heating exactly
balances the internal energy advection. If $\gamma < 5/3$,
the adiabatic heating alone cannot maintain $T \propto r^{-1}$
profile and viscous dissipation provides the additional heating
needed to maintain $r^{-1}$ profile.

To facilitate further discussions, we define five time scales
relevant in general accretion flow:
the inflow time scale
\begin{equation}
  t_{flow} \equiv r/v_r,
\end{equation}
the advective cooling time scale
\begin{equation}
  t_{adv} \equiv \frac{\varepsilon}{q^{-}_{adv}},
\end{equation}
the viscous heating time scale
\begin{equation}
  t_{vis} \equiv \frac{\varepsilon}{q^{+}_{vis}},
\end{equation}
the radiative heating time scale
\begin{equation}
  t_H \equiv \frac{\varepsilon}{H},
\end{equation}
and the radiative cooling time scale
\begin{equation}
  t_C \equiv \frac{\varepsilon}{C},
\end{equation}
where $r$ is the radius, $v_r$ the infall velocity,
$\varepsilon$ the internal energy of gas,
$q^{-}_{adv}$ the cooling rate due to advection,
${q^{+}_{vis}}$ the viscous heating rate,
$H$ the radiative heating rate, 
and $C$ the radiative cooling rate, all per unit volume.
In self-similar solutions, 
$t_{flow} = t_{adv}/\epsilon \simeq t_{adv}$ and
$t_{adv} = f^{-1} t_{vis} = (1-f)f^{-1} t_C$ 
where $f$ is the ratio between the advective cooling
and the total cooling rates and 
$\epsilon = (5/3-\gamma)/(\gamma-1)$; (NY).
When the flow is advection dominated, $f=1$ and
$t_{flow} \simeq t_{adv} \simeq t_{vis} \ll t_C$.

\section{Cooling in Two-Dimensional ADAF}

The self-similar ADAF is, like the Bondi solution,
very attractive due to its simplicity. 
Most physical quantities are simple power-laws of radius
and can be readily calculated. However, this also
implies that the solution may not be appropriate everywhere
for any real accretion flow. 

The simplicity of ADAF self-similar solution is based on
the simple energy balance equation.
The assumption of a constant ratio between
the viscous heating, advective cooling, and radiative cooling,
$1:f:1-f$, maintains the self-similarity of the solutions, yet
puts severe restrictions on cooling processes. Should the ratio
be constant in radius, the energy equation would be satisfied
at only one point. NY overcome this problem 
by finding a semi-self consistent
variable $f$ for applications to real accretion flows.
However, this is only possible when a positive $f$ can be found:
the cooling should be smaller than the heating. Solutions do not
exist, if positive $f$ is not possible. Cooling may be so strong
that the flow can not be kept at a high temperature (Rees et al. 1982).
This leads to the critical mass accretion $\dot M_{crit}$ 
for a given radius above which 
the optically thin hot solutions do not exist 
(Abramowicz et al. 1995; NY3). 
And for a given $\dot m$, hot solutions exist only within
some critical radius, $r_{crit}$. 

First, we will study similar issues in two-dimensional ADAF.
The viscous heating per unit volume in self-similar ADAF is simply
\begin{equation}\label{eq:qvis}
   q_{vis}^{+} = f^{-1} q_{adv}^{-} 
              = \case{3}{2}\epsilon^\prime \frac{\rho|v_r|c_s^2}{r},
\end{equation}
where $\rho(r,\vartheta)$ is the gas density, $v_r$ the radial
velocity, and $c_s$ the isothermal sound speed.
The composite formula for
atomic cooling and non-relativistic bremsstrahlung 
(Stellingwerf \& Buff 1982; Nobili, Turolla, \& Zampieri 1991) 
extended to relativistic bremsstrahlung is used for cooling,
\begin{equation}\label{eq:coolrate}
   C = \sigma_T c \alpha_f m_e c^2 n_i^2
   \Big[
   \big\{ \lambda_{br}(T_e)
          + 6.0\times10^{-22}\theta_e^{-1/2} \big\}^{-1}
         + \left(\frac{\theta_e}{4.82\times10^{-6}}\right)^{-12}
   \Big]^{-1} ,
\end{equation}
where $\sigma_T$ is the Thomson cross section,
$\alpha_f$ the fine-structure constant, $n_i$ the number
density of ions, and $\theta_e \equiv kT_e/m_ec^2$. 
The relativistic bremsstrahlung
rate is (Svensson 1982; Stepney \& Guilbert 1983; NY3)
\begin{equation}
   \lambda_{br} =  \left(\frac{n_e}{n_i}\right)(\sum_i Z_i^2)
                   F_{ei}(\theta_e) 
                 + \left(\frac{n_e}{n_i}\right)^2 F_{ee}(\theta_e),
\end{equation}
where
\begin{eqnarray}
   F_{ei} & = & 4 \left(\frac{2}{\pi^3}\right)^{1/2}
                \theta_e^{1/2}(1+1.781\theta_e^{1.34})
          \quad\hbox{for}~~ \theta_e < 1 \\
          & = & \frac{9}{2\pi} \theta_e
                \left[\ln(1.123\theta_e+0.48)+1.5\right]
          \quad\hbox{for}~~ \theta_e > 1 \nonumber\\
   F_{ee} & = & \frac{5}{6\pi^{3/2}}(44-3\pi^2)\theta_e^{3/2}
                (1+1.1\theta_e+\theta_e^2-1.25\theta_e^{5/2})
          \quad\hbox{for}~~ \theta_e < 1 \\
          & = & \frac{9}{\pi} \theta_e
                \left[\ln(1.123\theta_e)+1.2746\right]
          \quad\hbox{for}~~ \theta_e > 1 ,\nonumber
\end{eqnarray}
and $Z_i$ is the charge of ions.
The unknown electron temperature is assumed to be 
described by the fitting formula
$T_e = T_i T_a/(T_i+T_a)$ with $T_a \simeq 10^9 \:\hbox{K}$ and
$(kT_i/m_pc^2) = c_s^2(\vartheta)(r_s/r)^{-1}/4$ with
$r_s \equiv 2GM/c^2$, which
approximates the electron temperature profile in ADAF (NY3). 

The ratio $f$ is calculated from the definition,
\begin{equation}
   1 - f \equiv \frac{C}{q^+_{vis}}
\end{equation}
in the $(r,\vartheta)$ plane. 
The two-dimensional structure of the flow
is basically determined by one parameter 
$\epsilon^\prime \equiv \epsilon/f$ (see NY2).
Contours of $f=0.9$ (innermost curve), 
$0.8$, $0.7$, $0.6$, $0.5$, $0.3$, $0.2$, $0.1$, and $0.0$
(outermost curve) for ADAF with $\dot m=0.03$, 
$\epsilon^\prime \equiv \epsilon/f =1.0$,
and $\alpha=0.1$ are shown in Figure 1. 
The heavily shaded region above the contours
represents $f<0$, i.e.,
the region where hot electrons can not exist due to cooling.
The one-dimensional calculation of NY3 (Fig. 3)
shows that $f=0.5$ flow has $r_{crit} \sim 10^4 r_s$, 
which agrees well with $r_{crit} \simeq 1.1\times10^4 r_s$ 
for $\vartheta=0$ in Figure 1.

Note that there is a roughly
cone-shaped shaded region around the pole
where the hot solution is not possible.
For $\vartheta \ll 1$, this region extends down to $\sim r_s$.
This is mainly due to the slow
infall velocity near the pole (NY2) to which the advective cooling
is proportional. Since the viscous heating and radiative cooling are
proportional to the advective cooling in self-similar ADAF,
a small infall velocity near the pole implies a small advective cooling,
and hence small viscous heating and radiative cooling. But even
atomic plus bremsstrahlung cooling can have a far higher cooling rate than
is assumed in ADAF; therefore the flow cannot maintain the high
temperature in such regions. 
Normally, electrons in the region would cool down to 
$\sim 10^4\:\hbox{K}$. To first order,
the dynamics of the flow would not be affected as long
as the Coulomb coupling is weak and the ion temperature is near virial.
But the low temperature of the electrons for the gas near the polar axis
will make the coupling stronger 
[energy exchange rate $\propto T_e^{-1/2}$]
and it is possible that the ion temperature would consequently drop
below the virial value, resulting in the collapse of the polar region.
In fact, as we shall show in Park \& Ostriker (1999), this is the 
typical case.
The possibility of the collapse of the polar region of the flow
even at high temperature (for different reasons)
is also discussed by Blandford \& Begelman
(1998). This will create a funnel around the polar axis (Fig. 1)
and the flow will look more like torus than spheroid
(Rees et al. 1982; Paczy\'nski 1998).

\section{Comptonization}

When the accretion flow is quasi-spherical as in ADAF,
whatever photons are produced
at smaller radii inevitably interact with the outer part of the flow
on their way out.
In generic conditions, Compton scattering
of photons off electrons is usually the most important
interaction.
They can either heat or cool the gas depending on the
the spectrum of the radiation and the temperature of the gas.
High energy photons heat the electrons while low energy ones
cool them. However, the flow at large radius is generally
heated because most photons are produced at hotter inner regions,
and this is called preheating (Ostriker et al. 1976). 

The importance of preheating on the flow can be estimated by
comparing the relevant time scales. If 
$t_{vis} \ll t_H$, preheating can be safely ignored.
However, if there exists a region where 
$t_H < t_{vis}$, then preheating dominates over
the viscous heating or the advective cooling.
Since radiative fluxes are, in the lowest approximation,
quadratic in the flow rate (Shapiro 1973a,b), whereas advective
heating/cooling terms are linear in the flow rate, preheating
will become significant for high enough accretion rates.
In the spherical case we found that the solutions are significantly
altered for the mass accretion rate $\dot m \equiv 
\dot M / {\dot M}_{Edd} \gtrsim 10^{-1.5}$ and the luminosity 
$l \equiv L/L_{Edd} \gtrsim 10^{-7.5}$, 
with no high-temperature solutions possible having 
$l\gtrsim 0.03$ and $e\gtrsim 3\times10^{-4}$ 
due to the preheating instability (Park 1990a,b).
Since the density of the ADAF is higher than that in spherical
accretion flow for the same mass accretion rate due to the 
small infall velocity, we expect that this limit will occur
for higher values of $e$ and $l$. 

In ADAF, all physical quantities
are products of a radial part, a function of radius $r$ only, 
with an angular part, a function of spherical 
polar angle $\vartheta$ only. So 
\begin{equation}
  t_{vis} = \frac{1}{\epsilon^\prime} \Omega_K^{-1}(r) v^{-1}(\vartheta),
\end{equation}
where the radial infall velocity is defined as
$v_r = r \Omega_K(r) v(\vartheta)$
with $\Omega_K (r) \equiv (GM/r^3)^{1/2}$ and
\begin{equation}
  t_H = \left(\frac{m_e c^2}{4 k T_X}\right)
        \left(\frac{L_E}{L_X}\right)
        \case{3}{2} c_s^2 (\vartheta) \frac{r}{c} ,
\end{equation}
where $m_e$ is the electron mass, 
$c$ the speed of light,
$T_X$ the Compton
temperature of the radiation defined as 
the energy-weighted mean of photon energy averaged over the photon
spectral number density, $4kT_X \equiv <(h\nu)^2>/<h\nu>$,
$L_X$ the luminosity of the Comptonizing radiation,
$L_E$ the Eddington luminosity, $c_s(\vartheta)$ the
isothermal sound speed divided by the Keplerian velocity
$ c_s(r,\vartheta) \equiv (p/\rho)^{1/2} 
\equiv r \Omega_K(r) c_s(\vartheta) $,
$p$ the total pressure, and $\rho$ the gas density 
(Levich \& Syunyaev.1971).
The inequality $t_H < t_{vis}$ now reduces to
\begin{equation}\label{eq:tH_tvis}
  v(\vartheta) c_s^2(\vartheta) \left(\frac{r_s}{r}\right)^{1/2}
  < \case{2}{3}\left(\frac{4 k T_X}{m_e c^2}\right)
        \left(\frac{L_X}{L_E}\right).
\end{equation}
We take the case of the flow $L_X/L_E = 3\times10^{-4}$, 
$T_X = 10^9\:\hbox{K}$, and $\epsilon^\prime=0.1$ 
as an example (NY3). Equation (\ref{eq:tH_tvis})
represents the region above the solid curve in Figure 2,
in which preheating is the dominant heating process
and should not be ignored in the thermal balance equation.
In advection dominated flow,
$t_{flow} \simeq t_{adv} \simeq t_{vis}$ and, therefore,
$t_H < t_{flow}$. The flow is not adiabatic anymore
and dynamics of the flow will be significantly altered. 

Similarly, there could be many soft photons, thereby lowering
the radiation temperature $T_X \ll T_e$, and
the flow would be cooled by Compton scattering. The condition
for this is
\begin{equation}\label{eq:tH_tflow}
  v(\vartheta) c_s^2(\vartheta) \left(\frac{r_s}{r}\right)^{1/2}
  < \case{2}{3}\left(\frac{4 k T_e}{m_e c^2}\right)
        \left(\frac{L_X}{L_E}\right),
\end{equation}
which corresponds to the region above the dotted curve in Figure 2
for the same flow parameters.

For the flow near the equator, Compton heating or cooling may
be ignored within some radius. But above the equatorial plane
at large radius
or around the pole, Compton heating or cooling can be more important
than the viscous heating. It is quite possible that
Compton preheating would heat the flow near the pole
to a high temperature flow, that would otherwise cool down.
Or if there are abundant soft photons, the flow would cool
due to the Compton cooling. 
This result is independent of
the mass accretion rate and depends only on the Comptonizing
luminosity. Esin (1997) similarly found the importance
of Compton cooling (called `non-local cooling') under certain conditions
in a careful one-dimensional analysis, whereas Compton heating
was found to be generally unimportant. The main reason for this
discrepancy is that different radiation temperature, i.e., spectrum,
is assumed and that physical parameters of height-integrated
flow can be widely different from those of two-dimensional
flow, especially near the polar axis (NY2).

However, a real ADAF could be far more complicated than this.
Firstly, the radiation temperature $T_X$ can vary from place to place.
It should correctly represent the spectrum of the radiation
field at a given position, which is the sum of the radiation
transferred from other regions of the flow 
and that produced locally. 
Secondly, the luminosity $L_X$ is also a function of position, which is
again related to the density and temperature of the gas. 
Thirdly, the gas temperature is determined by the amount of Compton heating,
therefore the radiation temperature and the radiation energy
density. Hence, more accurate analysis requires simultaneously
solving the energy equations for gas and radiation.
This has been done only for spherical accretion flow
(Park 1990a,b; Nobili, Turolla, \& Zampieri 1991;
Mason \& Turolla 1992). 

\section{Preheating limit}

If preheating increases above some critical value,
it can affect the accretion flow more dramtically than
just changing the thermal balance. 
In Bondi-type spherical accretion flow, 
Ostriker et al. (1976) found that too much preheating changes
the dynamics of the flow around the sonic radius and would
disrupt the steady flow. This results in various
time-dependent behaviour in the flow and in the outcoming radiation
(Cowie, Ostriker, \& Stark 1978; Ciotti \& Ostriker 1997). 

Our next question is whether there is a similar preheating limit
for ADAF. Since ADAF is self-simliar, it does not
have an accretion radius or outer boundary. So we have to look
at the temperature structure in preheated ADAF. 
We will use simplified thermal balance equation to solve for
the temperature for the ADAF with strong preheating luminosity. 

In ADAF, the cooling rate is assumed to
be a constant fraction of the advective cooling to obtain
self-similar forms of solutions. This simplification may not
be too bad, if the solution is taken as height-integrated (or
angle-averaged) form. But in two-dimensional ADAF, the flow
time approaches infinity along the polar direction and 
is an order-of-magnitude larger than the usual 
free-fall time along the equatorial direction (NY2).
There is no radial advection along the pole, and 
we expect the Compton heating to be balanced by the radiative cooling.
Similarly we apply the thermal balance equation to all parts of
the flow to estimate the preheating effect, which is
valid when $t_H < t_{vis}$ (Fig. 2). 
The temperature is determined by requiring
$H(T_{eq};r,\vartheta) = C(T_{eq};r,\vartheta)$. 
When atomic line cooling and bremsstrahlung are the dominant processes,
$T_{eq}(r)$ has the form that outside some radius $r_\ast$,
$T(r) \simeq 10^4 \:\hbox{K}$ and at $r_\ast$, 
$T(r)$ jumps to $T_\ast \sim 2\times10^6 \:\hbox{K}$
at which temperature the bremsstrahlung cooling rate
becomes comparable to the peak of the atomic line cooling 
(Buff \& McCray 1974).
In this domain there is a classical phase change and
the temperature suddenly jumps,
because there is no stable equilibrium between $\sim 10^4 \:\hbox{K}$
and $T_\ast$. If we further assume that the
luminosity profile $L_X(r)$ is constant in $r$,
the temperature profile inside $r_\ast$ is simply 
$T(r) = T_\ast (r/r_\ast)^{-1}$ if the gas cools only by non-relativistic
bremsstrahlung.

At the transition radius $r_\ast$, Compton heating is 
equal to the peak of the cooling 
curve by definition (eq. \ref{eq:coolrate}),
\begin{equation}\label{eq:rstar}
  \frac{4k(T_X-T_e)}{m_ec^2} \frac{l}{\dot m} [n(\vartheta)]^{-1}
  \left(\frac{r_\ast}{r_s}\right)^{-1/2} = 8.8\times 10^{-5},
\end{equation}
where the gas number density is defined as
$n(r,\vartheta) = n(\vartheta)(\dot M/4\pi c m_p r_s^2)(r/r_s)^{-3/2}$.
Since the temperature of the electrons is not too different from
$\sim 10^9\:\hbox{K}$
at the inner part of the flow where most of the radiation 
is produced, we take $T_X \sim 10^9\:\hbox{K} \gg T_e(r)$. 
This choice of $T_X$
means that the radiation should contain enough hot photons
to heat the gas. If there are too many soft photons, e.g.,
synchrotron photons, $T_X$ could be lower.  

The accretion flow would be disrupted, if the flow is heated
above the virial temperature $T_{vir}$, defined as
$(\slantfrac{5}{2})kT_{vir} \equiv GM m_p /r$.
The flow temperature suddenly jumps
from $10^4\:\hbox{K}$ to $ T_\ast$ at radius $r_\ast$,
and the inflow would stop or reverse (i.e., become outflow) if
$T_\ast > T_{vir}(r_\ast)$.
The condition $T_\ast > T_{vir}(r_\ast)$ is equivalent to
$r_\ast > r_v \simeq 9.4 \times 10^5 r_s $ (eq. \ref{eq:rstar})
since $r_v$ is defined as $T_{vir}(r_v) = T_\ast$.

For the efficiency $e = l/\dot m = 3\times10^{-3}$ flow with
$\epsilon^\prime=1.0$ and $T_x = 10^9\:\hbox{K}$,
$r_\ast$ as a function of $\vartheta$ is shown in Figure 3 as a solid
curve, and $r_v$ as a dotted one. The part of the flow between
$r_\ast$ and $r_v$ with $r_\ast > r_v$ is overheated and the steady
inflow is not possible whereas the part of the flow with $r_\ast < r_v$
can accrete normally. The flow in regions A and C of Figure 3 has low
temperature $\sim 10^4\:\hbox{K}$ and is stable. The flow
in regions B and D are Compton heated above $T_\ast$ with
the flow in region B being unstable, while that in region D being stable.
So it would be possible that the flow
accretes along the equatorial plane while there is outflow
along the pole due to preheating.
Blandford \& Begelman (1998) also propose advection dominated
inflow-outflow via different mechanism: Part of the conservative
flow can have positive energy due to the energy flux transported
by the viscous torque (see also NY2). 

The critical efficiency $e_{cr}$ above
which this overheating starts to occur in any part of the flow
corresponds to $r_\ast|_{\vartheta=0} = r_v$
since preheating is most effective in the polar direction
due to lower infall velocity and lower density.
Figure 4 shows $e_{cr}$ determined for $T_X = 10^9\:\hbox{K}$
as squares.
Depending on the value of $\epsilon^\prime$, any flow
with efficiency above $\sim 4\times10^{-3}$ will suffer preheating instability
at or near the pole, and can develop the time-dependent behaviour
or outflow as in spherical case (Cowie, Ostriker, \& Stark 1978).
This value of $e_{cr} \sim 4\times10^{-3} $
corresponds to $l \simeq 2\times10^{-4}$
and $\dot{m} \simeq 0.05$ in NY3 solutions, 
and is comparable to that in the spherical accretion flow. 
Since the radiation temperature assumed
here is higher than that of the self-consistent
spherical flow (Park 1990a),  the critical efficiency, which
is inversely proportional to the radiation temperature,
should be smaller. 
However, the gas density of ADAF 
is on average $30 \sim 100 $ times greater for the same 
total mass accretion rate due to the smaller infall velocity
than that in the spherical accretion flow which is almost
freely falling (Park 1990a; NY2), resulting in similar critical
efficiency. For the same radiation temperature, i.e., spectral shape,
higher density provides a major advantage
for ADAF (in comparison with spherical flow): higher luminosities
and efficiencies should be possible.

\section{Luminosity from ADAF}

One of the unique difference between ADAF (or spherical accretion) 
onto black holes and thin disk accretion is that its radiation 
efficiency cannot be assumed a priori but must be determined
self consistently.
In thin disk accretion, whatever the energy input
to the flow, it is locally radiated away because of the long inflow time,
and its radiation efficiency is essentially determined by the
position of inner edge of the disk.
However, accretion flow with significant radial velocity can carry
the energy into the hole as well as radiates away. 
So the outcoming
radiation can be either significant or negligible depending on the dynamics
and thermal structure of the flow 
(see Park \& Ostriker 1998 for review). 

Here, we estimate how much outgoing radiation is produced 
self consistently by self-similar ADAF. 
The amount of radiative cooling in ADAF
is always assumed to be some fraction $(1-f)$ of the viscous heating.
The remaining fraction $f$ is advected with the flow and not radiated. 
Hence the radiative luminosity would be $(1-f)$ times 
the sum of all viscous heating (Eq. \ref{eq:qvis}) over
all radii and angles,
\begin{equation}\label{eq:Lrad}
   L_{rad} = \case{3}{2}\epsilon^\prime (1-f) \int_{r_{in}}^{r_{out}} dr
       \int_0^\pi 2\pi d\vartheta \rho|v_r|c_s^2 r.
\end{equation}
Since the parameter $f$ is not determined a priori for self-similar ADAF,
we take $f=0$ to estimate the maximum luminosity that can be produced.
Substituting physical quantities of NY2 
yields the dimensionless maximum luminosity
as a function of $\epsilon^\prime$. Since the right-hand side of
the equation (\ref{eq:Lrad}) is proportional to $\dot M$, the maximum
efficiency $e_{max}$ rather than the maximum luminosity is determined.
We assume $r_{in} = 3r_s$, and the values of $e_{max}$ for each
$\epsilon^\prime$ are shown in Figure 4 as circles. 
Comparison with the critical efficiency, $e_{cr}$, shows that 
the flow with $\epsilon^\prime \gtrsim 0.3$ has a possiblility of
preheating disruption. 
Higher $\epsilon^\prime$ flows are more vulnerable to preheating
because they have higher infall velocity and pressure,
therefore, higher viscous heating and radiation output.

The exact value of $e_{max}$ may differ somewhat from the values above,
because simple thermal balance equation
is not always valid. In reality, the local cooling rate can
be larger than the viscous heating because of the additional $PdV$
work. A very good example is the spherical accretion flow without
viscous dissipation. The viscous heating is zero, yet the gas is
heated by compression due to gravity and radiates
(Shapiro 1973a,b; Park 1990a,b). Therefore, $e_{max}$ can be higher.

\section{Self Consistent Flows}

So far we have discussed the cases where gas near the polar axis
or at large radius is cooled in the absence of preheating, or
it can be heated too much and disrupted by preheating. 
However, it is also possible
that the flow can be maintained at high temperature by preheating 
without being disrupted. In spherical accretion flow, there exist
two branches of solutions for certain mass accretion rate 
(Park 1990a,b; see Park \& Ostriker 1998 for references):
In the lower luminosity branch, gas is cooled down to 
$\sim 10^4 \:\hbox{K}$ and thus,
not much energy is released through radiation. 
In the other, higher luminosity branch,
gas at large radius is Compton heated by hot radiation
produced in the inner region, and a higher radiation efficiency is
achieved. We do find that this is also true for ADAF. For some
mass accretion rates, there exist hot solutions self-consistently
maintained by Compton preheating, whereas the flow would cool
down to thin disk in the absence of preheating (Park \& Ostriker 1999).

\section{Summary}

We have studied the thermal properties of 
the self-similar, two-dimensional advection dominated
accretion flow (NY2) with special consideration given to the radiative
cooling and heating. We find that

1. A hot solution is possible only within some critical radius 
for a given mass accretion rate in the equatorial plane, confirming
the one-dimensional analysis (NY3). Also, for any mass accretion
rate, a roughly conical region around the pole cannot maintain
high-temperature electrons in the absence of radiative heating
with the collapsed region shrinking as $\dot{m}$ is reduced.
If ions become coupled to the now cold electrons, (as seems likely)
an empty funnel around the polar axis will be created. 

2. Part of the flow at large radii or above the equatorial plane
should be affected through Compton heating or cooling by
photons produced at smaller radii, if the luminosity is high enough. 
If the radiation efficiency of the accretion is above $\sim 4\times10^{-3}$,
and the outcoming radiation has mean photon energy comparable to
the electron temperature of the inner region, preheating
due to the inverse Compton scattering would 
overheat the polar region of the flow,
and may create time-dependent behaviour or outflow while
accretion still goes on in the equatorial directions. For NY3
solutions, these phenomena should begin to occur for luminosities
$l \gtrsim 2\times10^{-4}$ and accretion rate $\dot{m} \gtrsim 0.05$
in Eddington units.

3. The role of Compton preheating is quite intriguing in ADAF as in
spherical flows. On the one hand it can have the attractive
feature of driving polar winds whenever the total luminosity
is above some rather low bound. But it may also allow another
branch of solutions making the original ADAF solutions more viable
for higher mass accretion rates than those for which 
solutions were valid. The detailed
calculation of this new branch of solutions will be given
in Park \& Ostriker (1999).

\acknowledgments

We would like to thankfully acknowledge useful conversations
with R. Narayan, I. Yi, X. Chen, B. Paczy\'nski, and R. Blandford.
This work is supported by NSF grant 
AST 9424416 and KOSEF 971-0203-013-2.
Large part of this work was done when MGP visited Princeton University
Observatory with the support from Professor Dispatchment Program
of Korea Research Foundation and NSF 9424416.

\appendix

\clearpage

\begin{figure}
\plotone{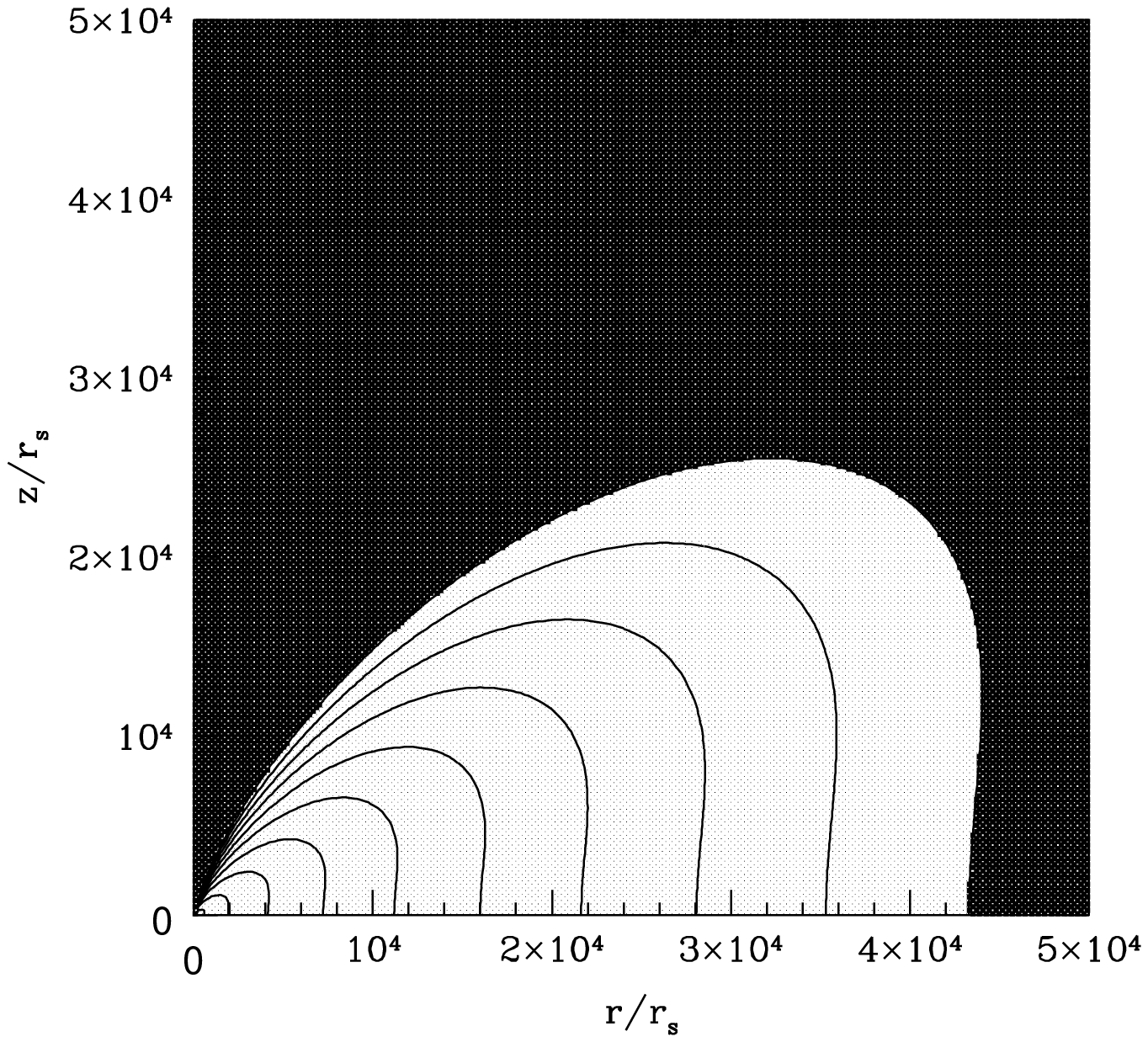}
\caption{Contours of $f=0.9$ (innermost curve), 
$0.8$, $0.7$, $0.6$, $0.5$, $0.3$, $0.2$, $0.1$, and $0.0$
(outermost curve) for $\dot m=0.03$, 
$\epsilon^\prime \equiv \epsilon/f =1.0$,
and $\alpha=0.1$. Heavily shaded region above the contours 
has $f<0$, i.e., hot electrons
can not exist due to excessive cooling.}
\end{figure}

\clearpage

\begin{figure}
\plotone{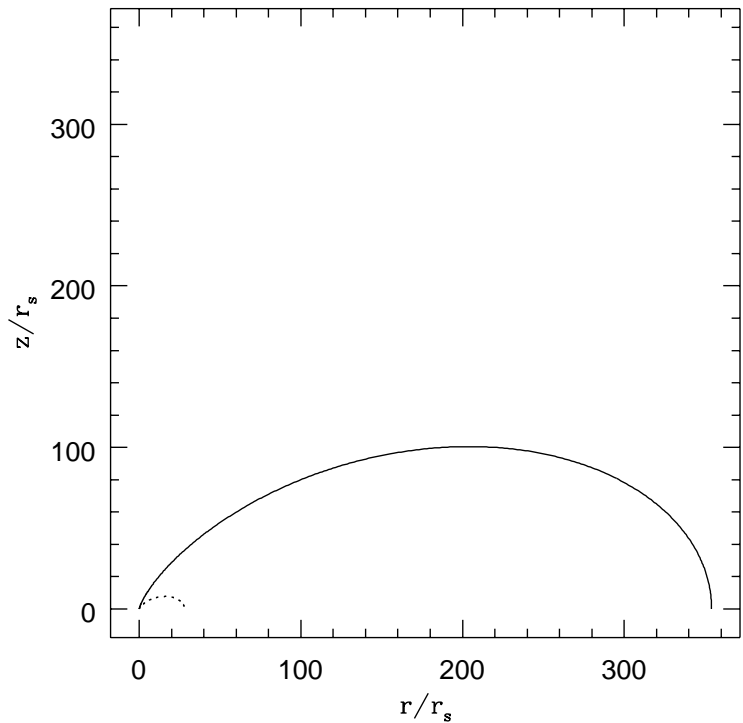}
\caption{In the region above the solid curve the Compton heating
         time scale is shorter than the viscous heating time scale and,
         approximately, the inflow time scale
         for $L_X/L_E = 3\times10^{-4}$, $T_X = 10^9\:\hbox{K}$,
         and $\epsilon^\prime=0.1$ flow. Above the small dotted curve,
         the Compton cooling time scale is shorter than the viscous
         time scale if $T_X \ll T_e$. Radius $r$ and disk height $z$
	 are in units of Schwarzschild radius $r_s$.}
\end{figure}

\begin{figure}
\plotone{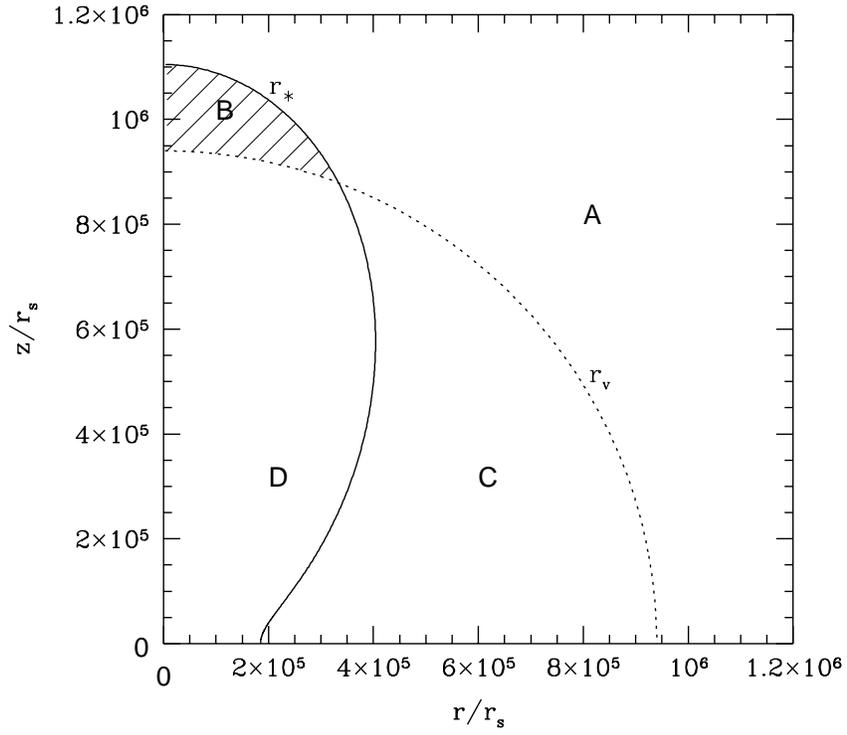}
\caption{For $e = 3\times10^{-3}$ flow with
         $\epsilon^\prime=1.0$ and $T_x = 10^9\:\hbox{K}$,
         $r_\ast$ is shown as a solid
         curve, and $r_v$ as a dotted one.
         In regions A and C the flow is cool ($T\simeq10^4\:\hbox{K}$);
         while in B and D it is hot ($T>T_\ast$). 
         In region B the temperature is above the virial temperature,
         the energy is positive and infall would be impossible.}
\end{figure}

\begin{figure}
\plotone{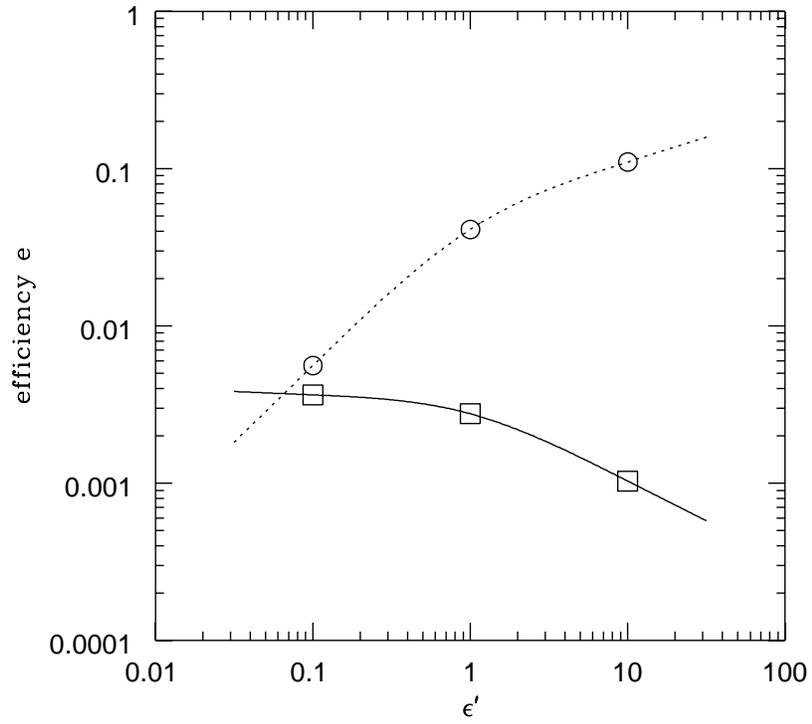}
\caption{The critical efficiency $e_{cr}$ (squares)
         at which overheating starts to occur for $T_X = 10^9\:\hbox{K}$
         as a function of the parameter $\epsilon^\prime$.
         The maximum radiation efficiency $e_{max}$ (circles) 
         by assuming all viscous heating being converted to radiation.}
\end{figure}

\end{document}